**Tunable optoelectronic and ferroelectric properties in Sc-based III-nitrides**

Siyuan Zhang[1], David Holec[2], Wai Y. Fu[1], Colin J. Humphreys[1], Michelle A. Moram[1,3]

*1. Department of Materials Science and Metallurgy, University of Cambridge, Pembroke Street, CB2 3QZ, Cambridge, UK*

*2. Department of Physical Metallurgy and Materials Testing, Montanuniversität Leoben, Franz-Josef-Straße 18, 8700 Leoben, Austria*

*3. Department of Materials, Imperial College London, Exhibition Road, SW7 2AZ, London, UK*

**Abstract**

Sc-based III-nitride alloys were studied using Density Functional Theory with special quasi-random structures and were found to retain wide band gaps which stay direct up to $x = 0.125$ (Sc$_x$Al$_{1-x}$N) and $x = 0.375$ (Sc$_x$Ga$_{1-x}$N). Epitaxial strain stabilization prevents spinodal decomposition up to $x = 0.3$ (Sc$_x$Al$_{1-x}$N on GaN) and $x = 0.24$ (Sc$_x$Ga$_{1-x}$N on GaN), with critical thicknesses for strain relaxation ranging from 3 nm to near-infinity. The increase in Sc content introduces compressive in-plane stress with respect to AlN and GaN, and leads to composition- and stress-tunable band gaps and polarization, and ultimately introduces ferroelectric functionality in Sc$_x$Ga$_{1-x}$N at $x \approx 0.625$.

**Main Article**

The wurtzite-structure III-nitride semiconductors AlN, GaN, InN, and their alloys are used widely in short wavelengths light emitting diodes, lasers and in high electron mobility transistors (HEMT). However, lattice and polarization mismatches between layers of different compositions fundamentally limit the range of possible device designs [1,2].

Therefore, we explored the effects of alloying ScN with AlN and GaN. ScN is a semiconductor with a direct band gap of 2.1 eV and an indirect band gap of 0.9 eV in the ground state rocksalt structure [3-6]. Additionally, ScN is predicted to be metastable in the hexagonal *h*-BN phase [7]. This non-polar structure is predicted to have remarkable strain-tunable optical and piezoelectric properties and can be derived from the wurtzite structure as the *c/a* lattice parameter ratio reduces from ~1.6 to ~1.2 and as the internal parameter *u* simultaneously increases from ~0.38 to 0.5 [8]. $Sc_xAl_{1-x}N$ and $Sc_xGa_{1-x}N$ alloys might therefore retain the hexagonal structures and wide band gaps needed for combination with other III-nitrides for optoelectronic and electronic device applications. $Sc_xAl_{1-x}N$ and $Sc_xGa_{1-x}N$ alloys may also exhibit a wide, tunable range of properties if they can be stabilized in structures varying from wurtzite-like to *h*-BN-like. Indeed, a characteristic reduction in the *c/a* ratio of relatively thick wurtzite-like $Sc_xAl_{1-x}N$ [9,10] and $Sc_xAl_{1-x}N$ films [11-14] has already been observed experimentally for low Sc concentrations. However, two major limitations have hindered the exploration of tunable properties within these systems. (i) Previous theoretical calculations have artificially constrained these alloys into the wurtzite structure (with fixed *c/a*) [15], have introduced artificial ordering of the metal atom positions [15,16], which is not observed experimentally, and/or have not calculated properties of interest for electronic devices [17] over the full alloy composition range [12]. (ii) The extent to which epitaxial strain engineering could stabilize $Sc_xAl_{1-x}N$ and $Sc_xGa_{1-x}N$ alloys thermodynamically and/or further tune their properties remains unknown. These effects are important: for example, strain stabilization allows $In_xGa_{1-x}N$ quantum wells to be grown homogeneously [18], as needed for solid state lighting. In this work, we explore the stability of $Sc_xAl_{1-}$

$_x$N and Sc$_x$Ga$_{1-x}$N alloys, assess the tunability of their properties and determine the true potential of these materials for use in multifunctional III-nitride-based devices.

The structural and electronic properties of Sc$_x$Al$_{1-x}$N and Sc$_x$Ga$_{1-x}$N alloys were calculated using Density Functional Theory (DFT) [19] and the special quasi-random structure (SQS) [20,21] methodology to best represent a random alloy. The Vienna Ab-initio Simulation Package (VASP) [22] was used for structural optimization, using the general gradient approximation (GGA) [23] for the exchange-correlation potential. The lattice parameters and internal coordinates of the structures were allowed to relax fully at each composition to achieve the lowest energy configuration. Epitaxially strained hexagonal supercells have fixed *a*, while other parameters were fully relaxed. The structural optimization reaches energy accuracy of < 1meV/atom. The WIEN2k code [24] was used for band structure calculations, with the modified Becke-Johnson (mBJ) potential [25] implemented in WIEN2k to counteract the underestimation of band gaps which usually occurs in DFT calculations. The details of our calculation methods are given in the Supplemental Material [26].

Fig. 1 shows the relative stability of different Sc$_x$Al$_{1-x}$N and Sc$_x$Ga$_{1-x}$N structures by examining their mixing enthalpies with respect to the ground state phases, rocksalt ScN and wurtzite AlN, GaN. Hexagonal structures are favored at lower Sc contents, whereas rocksalt structures are favored at higher Sc contents, consistent with previous experiments [10,11,13] and calculations [15,17]. Alloy ordering parallel and perpendicular to the (0001) plane was also investigated, but the reduction in mixing enthalpies was never larger than the entropy term at a typical molecular beam epitaxy (MBE) growth temperature of 1100 K [26]. Therefore, ordered hexagonal Sc$_x$Al$_{1-x}$N and Sc$_x$Al$_{1-x}$N alloys are unlikely to form in practice, consistent with experiments [10,11,13].

Fig. 2 indicates that the *c/a* ratios of the relaxed hexagonal structures decrease with increasing Sc content, consistent with trends found for experimental data [9,11] and in agreement with previous calculations [17]. Within the stable range of Sc$_x$Al$_{1-x}$N for the hexagonal structures, the *c/a* ratios decrease from 1.60 (AlN) to ~1.5 at the highest stable Sc content of *x* = 0.5, where the structure remains wurtzite-like. However, within the stable range of $Sc_xGa_{1-x}N$ for the hexagonal structures, the $c/a$ ratios decrease from 1.63 (GaN) to ~1.25 at the highest stable Sc content of $x = 0.625$, where the structure is $h$-BN-like and almost non-polar ($u = 0.494$). The important difference between $Sc_xAl_{1-x}N$ and $Sc_xGa_{1-x}N$ alloys is highlighted in Fig. 1, namely the hexagonal polar to non-polar (wurtzite-like to $h$-BN-like) phase transition of $Sc_xGa_{1-x}N$ between $x = 0.5$ and 0.625, whereas the $h$-BN-like structure is only metastable for $Sc_xAl_{1-x}N$ alloys [17].

As the high mixing enthalpies of $Sc_xAl_{1-x}N$ and $Sc_xGa_{1-x}N$ alloys indicate (Fig. 1), our thermodynamic calculation of the alloys shows that $Sc_xAl_{1-x}N$ and $Sc_xGa_{1-x}N$ are only stable with respect to spinodal decomposition up to $x = 0.045$ and $x = 0.09$ at 1100 K, respectively (Fig. 3) [26]. Indeed, phase separation has been observed experimentally in $Sc_xAl_{1-x}N$ [9,10] and $Sc_xGa_{1-x}N$ [11]. On one hand, the effect of epitaxial strains on their mixing enthalpies and spinodal decomposition temperatures was studied, up to a Sc content of $x = 0.375$. Fig. 3 shows the substantial stabilizing effect of epitaxial strains at 1100 K: $Sc_xAl_{1-x}N$ would be stabilized up to $x = 0.18$ if epitaxial strained to AlN, or up to $x = 0.30$ if strained to GaN; $Sc_xGa_{1-x}N$ would be stabilized up to $x = 0.24$ if epitaxial strained to GaN [26]. On the other hand, the use of lower growth temperatures could also kinetically limit the diffusion-controlled spinodal decomposition process for higher Sc contents. This appears to have occurred experimentally in the low-temperature growth of sputtered $Sc_xAl_{1-x}N$ [10].

Although epitaxial strain can stabilize hexagonal $Sc_xAl_{1-x}N$ and $Sc_xGa_{1-x}N$, it can only be maintained up to a certain critical thickness before strain relaxation occurs, typically by dislocation glide. We therefore calculated the equilibrium critical thicknesses using an energy balanced equilibrium model validated on wurtzite $In_xGa_{1-x}N$ and $Al_xGa_{1-x}N$ alloys [27,28]. As shown in Fig. 4, the calculated critical thicknesses of stable wurtzite-like $Sc_xAl_{1-x}N$ and $Sc_xGa_{1-x}N$ alloys lie between 3 nm (for mismatched $Sc_xAl_{1-x}N$ on AlN) to full lattice-matching (for $Sc_xAl_{1-x}N$ on GaN), making them of practical use as quantum wells for optoelectronics and in HEMT applications. Epitaxial strains were found to have minimal effects on alloy band gaps [26].

Fig. 5a shows band structures calculated for wurtzite AlN, GaN, ScN (unstable phase by imposing *c/a* = 1.62), and wurtzite-like $Sc_xAl_{1-x}N$ and $Sc_xGa_{1-x}N$ alloys. The band structures are aligned according to the charge neutrality level (CNL), which is estimated from the average of integrated eigenvalues of the top valence band and the bottom conduction band [29]. The alloy band structures both vary smoothly from the pronounced conduction band minimum (CBM) and valence band maximum (VBM) found at the Γ point in AlN and GaN, to the remarkably flat conduction and valence bands of wurtzite ScN due to the spatially localized Sc-3*d* states. However, the trends of band-gap values for $Sc_xAl_{1-x}N$ and $Sc_xGa_{1-x}N$ are different.

Fig. 5b shows that the CBM of $Sc_xAl_{1-x}N$ is dominated by Sc-3*d* states. Therefore, Sc-N hybridization determines the band gaps of $Sc_xAl_{1-x}N$ alloys which decrease as the Sc content increases. This trend is expected because the calculated band gap of wurtzite ScN is substantially lower than the AlN band gap (Fig. 6). The $Sc_xAl_{1-x}N$ band gaps remain direct (Γ→Γ) up to $x = 0.125$, above which the CBM and VBM moves from Γ towards M and the gap is no longer direct (Fig. 5a). However, the maximum difference between direct and indirect band gaps does not exceed 0.1 eV for $Sc_xAl_{1-x}N$ at $x \leq 0.5$ [26], and therefore direct behavior may still dominate experimentally.

Fig. 5c shows that the CBM of $Sc_xGa_{1-x}N$ is occupied by states from Ga and N, whereas the Sc-3*d* states are not present at the band edge. Therefore, Ga-N hybridization determines the band gaps of $Sc_xGa_{1-x}N$ alloys which increase as the Sc content increases, as is expected due to the larger band gap of wurtzite ScN than GaN (Fig. 6). The $Sc_xGa_{1-x}N$ band gaps remain direct (Γ→Γ) up to $x = 0.375$, above which the maximum difference between direct and indirect band gaps is still smaller than 0.01 eV for $Sc_xGa_{1-x}N$ at $x \leq 0.5$ [26], which suggests that direct behavior is very likely to dominate experimentally.

However, our results of $Sc_xGa_{1-x}N$ band gaps differ from previous calculations which predicted monotonous decrease in $Sc_xGa_{1-x}N$ band gaps with increasing Sc content [15], because (i) our structures have been allowed to relax to achieve the lowest energy configuration and (ii) we have not introduced artificial ordering into the unit cell. Neglecting these two factors led to unstable structures with band gaps very close to those published in Ref. [15]. Moreover, we considered ordering but fully relaxed the supercells, and still found monotonous increase in $Sc_xGa_{1-x}N$ band gaps with increasing Sc content up to $x = 0.5$ [26], confirming that improper relaxation of the supercell structures are responsible for the previous results [15].

Interestingly, previous optical absorption experiments on $Sc_xGa_{1-x}N$ films suggested a reduction in band gap as the Sc content increased [11]. Our predictions indicate that phase separation by spinodal decomposition should occur for strain-relaxed $Sc_xGa_{1-x}N$ films of $x > 0.09$ at reported growth temperatures of ~1070 K [11,12]. This could produce nanoscale regions with rocksalt-like crystal structures, potentially leading to lower energy optical absorption onsets as observed experimentally [11,12]. Additionally, an impurity or defect band within the gap is also likely to be present in the previous samples studied, as all contain a very high density of stacking faults [11,12]. In contrast, photoluminescence excitation experiments, performed on our own high-purity and relatively low defect density $Sc_xGa_{1-x}N$ films with $x \approx 0.06$ grown by MBE [13,14], indicate band gaps greater than that of the GaN substrate, consistent with our DFT predictions for this composition.

As the Sc content increases, both $Sc_xAl_{1-x}N$ and $Sc_xGa_{1-x}N$ go through a transition in which the $c/a$ ratio drops from ~1.6 to ~1.2 (Fig. 2). This is driven by the energetic stabilization associated with the transition from $sp^3$ to $sp^3d$ hybridization around the Sc atoms. After this transition, hexagonal $Sc_xAl_{1-x}N$ alloys are no longer structurally stable and instead prefer the rocksalt phase. In contrast, $Sc_xGa_{1-x}N$ alloys up to $x = 0.625$ are stable in the $h$-BN-like configuration (Fig. 1). The polar to non-polar phase transition of $Sc_xGa_{1-x}N$ happens between $x = 0.5$ ($u = 0.390$) and $x = 0.625$ ($u = 0.494$) (Fig. 2), and is characterized by the wurtzite $u$. The $u$ parameter can not only be tuned by compositions, but also by in-plane strains, as shown in Fig. 7. Between in-plane lattice parameters of 3.5 Å and 3.6 Å, $u$ can be tuned between 0.4 and 0.5. Although this has been demonstrated in the system of $h$-BN phase ScN, appreciable strains are required to induce the phase transition (strain energy 0.08~0.21 eV/formula unit), so is the case of $Sc_{0.5}Ga_{0.5}N$ (strain energy ~0.06 eV/formula unit). On the other hand,

$Sc_{0.625}Ga_{0.375}N$ shows exceptionally low strain energy (< 0.01 eV/formula unit, Fig. 7) in the tuned $u$ regime. The flat strain energy curve resembles the potential energy landscape of $Sc_{0.5}Al_{0.5}N$ [17], which is responsible for the high piezoelectric response of $Sc_xAl_{1-x}N$ alloys.

The possibility of ferroelectric polarity switching in $Sc_{0.625}Ga_{0.375}N$ along the [0001] axis was also investigated, and the magnitude and barrier energy of polarity switching at various in-plane strains are plotted in Fig. 7. The barrier energy for ferroelectric switching was determined by the energy difference between the strained supercell with its equilibrium $u$ parameter and with the non-polar case of $u = 0.5$, the saddle point for switching. For example, as can be read from Fig. 7, if a $Sc_{0.625}Ga_{0.375}N$ film is grown on top of an InN substrate ($a = 3.544$ Å), the equilibrium $u$ will be ~0.45, and the barrier energy will be ~0.05 eV/formula unit. This compares favorably to that of conventional ferroelectrics such as $PbTiO_3$ (0.1 eV/formula unit) and $BaTiO_3$ (0.02 eV/formula unit) [32]. For comparison, the switching barrier was also calculated for GaN and was found to be 0.60 eV/formula unit, corresponding to switching fields greater than the dielectric breakdown limit of ~$5\times10^6$ Vcm$^{-1}$, indicating no ferroelectric behavior and consistent with experiment. The ferroelectric switching barrier in the $Sc_xGa_{1-x}N$ alloys was found to be directly related to $u$ and independent of composition, but the strain energy associated with switching was lowest for $x = 0.625$. The ease of switching indicates that both ferroelectric functionality and high piezoelectric coefficients could be realized and tuned in $Sc_{0.625}Ga_{0.375}N$ films by strain engineering on different substrates. Spinodal decomposition of such alloys could be limited by the use of low growth temperatures in practice.

In summary, we have calculated that hexagonal $Sc_xAl_{1-x}N$ and $Sc_xGa_{1-x}N$ alloys meet important criteria for practical use in III-nitride-based devices. Their advantages include (i) wide band gaps which remain fully direct up to $x = 0.125$ for $Sc_xAl_{1-x}N$ and $x = 0.375$ for $Sc_xGa_{1-x}N$, (ii) the possibility of epitaxial phase stabilization of wurtzite-like $Sc_xAl_{1-x}N$ and $Sc_xGa_{1-x}N$ phases by growth on AlN and GaN, with critical thicknesses ranging from 3 nm to near-infinity for lattice-matched $Sc_xAl_{1-x}N$ on GaN, (iii) the new ability to introduce compressive in-plane stress by $Sc_xAl_{1-x}N$ alloys for control of wafer curvature and cracking within $Al_xGa_{1-x}N$-based ultraviolet light-emitting devices (Fig. 6), (iv) the ability to introduce lattice-matched barriers using $Sc_xGa_{1-x}N$ alloys in $In_xGa_{1-x}N$-based green light-emitting devices (Fig. 6), (v) the possibility for composition- and stress-tunable polarization for polarization-matching by $Sc_xAl_yGa_{1-x-y}N$ alloys [33] for use in ultraviolet light-emitting devices and HEMTs, and (vi) tunable ferroelectric and piezoelectric functionality for $Sc_xGa_{1-x}N$ around $x \approx 0.625$. These properties, tunable by composition and epitaxial strain, indicate valuable future applications for $Sc_xAl_{1-x}N$ and $Sc_xGa_{1-x}N$ alloys.

MAM acknowledges support from the Royal Society through a University Research Fellowship. DH acknowledges financial support from the FWF Start Project (Y371).

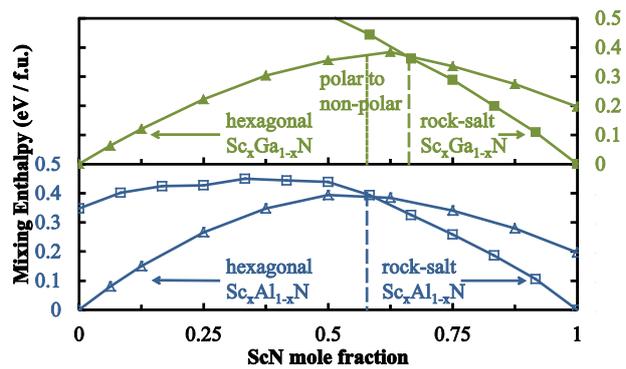

FIG. 1. Mixing enthalpies of $Sc_xAl_{1-x}N$ and $Sc_xGa_{1-x}N$ alloys in hexagonal and rocksalt phases. Lines are guides to the eyes.

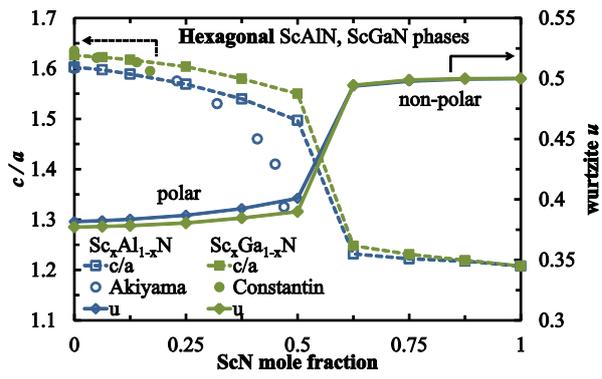

FIG. 2. Unit cell *c/a* ratios of hexagonal $Sc_xAl_{1-x}N$ and $Sc_xGa_{1-x}N$ alloys. Experimental data points for $Sc_xAl_{1-x}N$ and $Sc_xGa_{1-x}N$ *c/a* are adapted from [9] and [11], respectively. Lines are guides to the eyes.

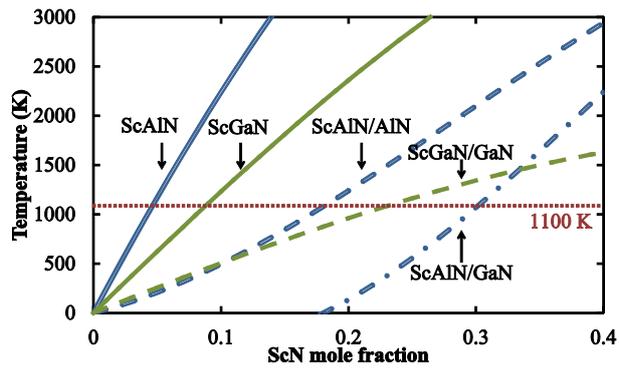

FIG. 3. The variation of spinodal decomposition temperatures of hexagonal $Sc_xAl_{1-x}N$ and $Sc_xGa_{1-x}N$ alloys with Sc content, relaxed or strained to the nearest III-nitride binary compounds (see [26] for full calculation details).

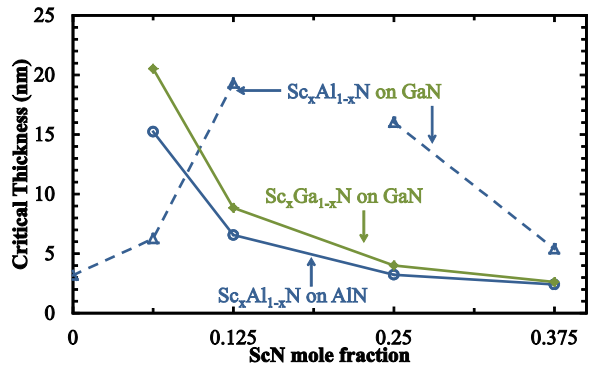

FIG. 4. Critical thicknesses for strain relaxation by dislocation glide of hexagonal $Sc_xAl_{1-x}N$ and $Sc_xGa_{1-x}N$ alloys on AlN or GaN considered for the $\frac{1}{3}<11\bar{2}3>\{1\bar{1}01\}$ slip system [27,28]. Lines are guides to the eyes.

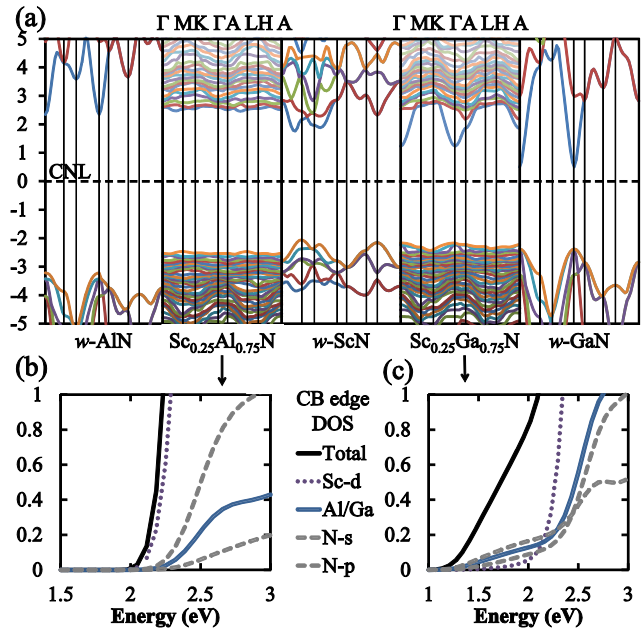

FIG. 5. (a) Band structures of wurtzite AlN, GaN, ScN, $Sc_xAl_{1-x}N$ and $Sc_xGa_{1-x}N$ alloys (x = 0.25). Projected density of states (DOS) at conduction band (CB) edges of wurtzite (b) $Sc_{0.25}Al_{0.75}N$ and (c) $Sc_{0.25}Ga_{0.75}N$ alloys. The energy 0 is the CNL of each system. Gaussian broadening of 0.14 eV was applied to DOS curves.

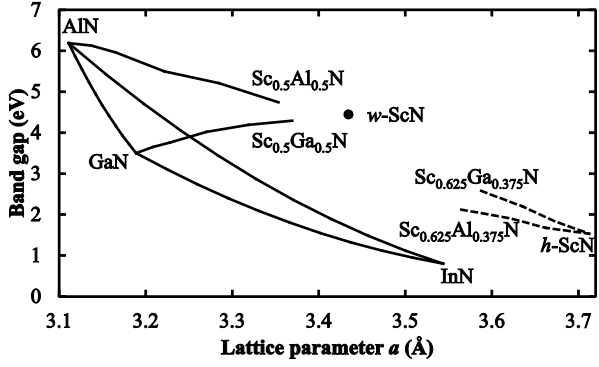

FIG. 6. Band gaps of AlGaN, AlInN, GaInN alloys, hexagonal $Sc_xAl_{1-x}N$ and $Sc_xGa_{1-x}N$ alloys as functions of lattice parameter $a$ (wurtzite-like structures in solid lines, $h$-BN-like structures in dashed lines). Band gaps of wurtzite and $h$-BN ScN phases are added for reference. Lattice parameters of $Sc_xAl_{1-x}N$ and $Sc_xGa_{1-x}N$ alloys were adapted from our GGA calculations corrected to match experimental lattice parameters of AlN and GaN [26]. Lattice parameters of other III-nitride alloys were estimated from Ref. [30]. Band gaps of $Sc_xAl_{1-x}N$ and $Sc_xGa_{1-x}N$ alloys were adapted from our mBJ-GGA calculations with a rigid upward shift of 0.62 eV for $x \leq 0.5$ to match the experimental band gaps of AlN and GaN. Band gaps of other III-nitride alloys were estimated by the bowing parameters suggested in Ref. [31].

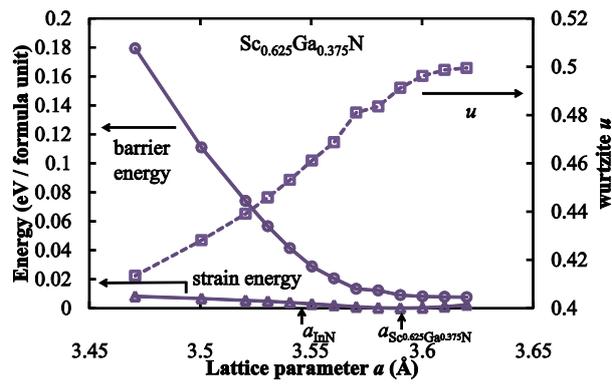

FIG. 7. The magnitude (proportional to the wurtzite $u$ parameter) and barrier energy for polarity switching of hexagonal $Sc_{0.625}Ga_{0.375}N$ alloy as a function of the lattice parameter $a$. Lines are guides to the eyes.